\documentclass[aps,twocolumn,showkeys,amsmath,amssymb,10pt,superscriptaddress]{revtex4}

\usepackage[colorlinks=true,citecolor=blue,anchorcolor=red,menucolor=red, linkcolor=red,filecolor=red,runcolor=red,urlcolor=blue,frenchlinks=true]{hyperref}
\usepackage{graphicx}
\usepackage{dcolumn}
\usepackage{txfonts}
\usepackage{bm}

\usepackage{amsfonts}
\usepackage{amsmath}
\usepackage{subfigure}
\usepackage{booktabs}
\usepackage{inputenc}
\usepackage{graphics}
\usepackage{dsfont}
\usepackage{mathtools}
\usepackage{autobreak}
\usepackage{diagbox}
\usepackage{multirow}

\usepackage{longtable}
\usepackage{etoolbox}
\apptocmd{\sloppy}{\hbadness 10000\relax}{}{}

\usepackage{overpic}
\usepackage{indentfirst}
\usepackage{cancel} 
\usepackage{slashed}  
\usepackage{cases}
\usepackage{color}
\usepackage{multirow}
\usepackage{natbib}
\allowdisplaybreaks[4]
\newcommand{\qq}{\langle\bar qq\rangle}

\newcommand{\GGa}{\langle GG\rangle}

\newcommand{\qGqb}{\langle\bar{q} Gq\rangle}

\newcommand{\GGGb}{\langle g_s^3G^3\rangle}

\begin{document}
\title{Subtraction of infrared divergences in light-quark QCD sum rules}
\author{Ding-Kun Lian}
\email{liandk@seu.edu.cn}
\affiliation{School of Physics, Southeast University, Nanjing 210094, China}
\author{Jin-Peng Zhang}
\affiliation{School of Physics, Sun Yat-Sen University, Guangzhou 510275, China}
\author{Zi-Xi Ou-Yang}
\affiliation{School of Physical Sciences, University of Chinese Academy of Sciences, Beijing 100049, China}
\author{Qi-Nan Wang}
\email{wangqinan@bhu.edu.cn}
\affiliation{College of Physical Science and Technology, Bohai University, Jinzhou 121013, China}
\author{Hua-Xing Chen}
\email{hxchen@seu.edu.cn}
\affiliation{School of Physics, Southeast University, Nanjing 210094, China}
\author{Wei Chen}
\email{chenwei29@mail.sysu.edu.cn}
\affiliation{School of Physics, Sun Yat-Sen University, Guangzhou 510275, China}
\affiliation{Southern Center for Nuclear-Science Theory (SCNT), Institute of Modern Physics, Chinese Academy of Sciences, Huizhou 516000, Guangdong Province, China}
\author{Jia-Jun Wu}
\email{wujiajun@ucas.ac.cn}
\affiliation{School of Physical Sciences, University of Chinese Academy of Sciences, Beijing 100049, China}
\affiliation{Southern Center for Nuclear-Science Theory (SCNT), Institute of Modern Physics, Chinese Academy of Sciences, Huizhou 516000, Guangdong Province, China}

\begin{abstract}
    In QCD sum rules for light-quark systems, infrared (IR) divergences can appear in the Wilson coefficients of certain condensates. These divergences manifest explicitly in the coordinate-space expressions of the light-quark propagators. We propose an improved method to eliminate these IR divergences at the propagator level and present a subtraction formula that implements this procedure. Compared to the existing methods that rely on the mixing between quark and gluon condensates of the same dimension to eliminate IR divergences, this method is more intuitive and easier to apply in practical QCD sum rule calculations. 
\end{abstract}
\maketitle

\section{Introduction}
The QCD sum rule approach, proposed by Shifman, Vainshtein and Zakharov~\cite{Shifman:1978bx,Shifman:1978by}, provides a powerful framework for studying hadron properties~\cite{Reinders:1984sr,Colangelo:2000,Gubler:2018ctz}. It utilizes the operator product expansion (OPE) to separate the short-distance effects, which can be calculated perturbatively, from the long-distance effects, which are parameterized by the vacuum condensates, such as the quark condensate $\qq$ and the gluon condensate $\GGa$. However, in light-quark systems, infrared (IR) divergences may arise in the Wilson coefficients of certain condensates~\cite{Gubler:2009ay,Huang:2014hya,Bandyopadhyay:2016inp,Li:2024rrs,Li:2025dkw,Zhang:2025fuz,Wang:2024lnv}. These divergences are due to the massless limit of the light quark, which leads to the absence of an IR regulator for the low-momentum behavior of the light-quark propagator.

Two approaches were proposed to address this issue. The first method, known as the heavy-quark expansion technique~\cite{Shifman:1978bx,Generalis:1983hb,Bagan:1985zp,Grozin:1986xh,Grozin:1994hd}, retains a finite light-quark mass $m_{q}$ throughout the calculation and expands the bilinear quark condensates in terms of the gluon condensates of the same dimension. The expanded contributions are then subtracted from the corresponding IR-divergent Wilson coefficients of the gluon condensate terms, canceling the divergences, and the limit $m_{q}\rightarrow 0$ is taken at the final stage. The second method, which offers greater computational simplicity, sets $m_{q}=0$ {\it ab initio} and employs dimensional regularization in the calculation~\cite{Broadhurst:1984rr}. Crucially, both methods rely on mixing between quark and gluon condensates of the same dimension, which produces divergent mixing coefficients that are then used to cancel the IR divergences in the Wilson coefficients. These methods have been successfully applied to remove the IR divergences in dimension-eight gluon condensates~\cite{Broadhurst:1985js,Huang:2014hya}. Notably, for the tri-gluon condensate $\GGGb$, IR divergences disappear when all corresponding diagrams are taken into account, making mixing with other condensates unnecessary, as shown in Refs.~\cite{Li:2024rrs,Li:2025dkw,Zhang:2025fuz}.

However, the elimination of IR divergences in the two approaches is implemented at the OPE level, meaning that one must first complete the OPE calculation and then perform the subtraction to remove the IR divergences. This renders the procedure more cumbersome. In this paper, we propose an improved subtraction method to eliminate IR divergences at the propagator level within dimensional regularization. In this method, inserting the interaction terms into the background quark propagator generates a one-loop contribution that is expressed in terms of the background gluon fields and serves as an IR counterterm. Subtracting this counterterm from the corresponding IR-divergent quark propagator yields an IR-finite quark propagator. This approach is more straightforward and computationally easier for practical QCD sum rule calculations involving light quarks. 

The paper is organized as follows. In Section~\ref{sec:2}, we show the IR divergences of the light-quark propagators in coordinate space, which are due to the massless limit for the light quarks. Section~\ref{sec:3} presents our method for removing these divergences at the propagator level and demonstrates its importance for obtaining the correct spectral density. Finally, the discussion and conclusions are provided in Section~\ref{sec:4}.

\section{IR divergences in light-quark propagators}\label{sec:2}
In light-quark QCD sum rules, it is convenient to work in the coordinate space when the light-quark propagators carry low-dimensional operators. For instance, the expressions for the perturbative propagator (corresponding to the unit operator) and the propagator with a background gluon field strength tensor in $D$ dimensions are comparatively compact:
\begin{align}
    S_{ab}^{pert}(x)=&\frac{i}{2\pi^{D/2}}\Gamma(\frac{D}{2})\frac{\slashed{x}}{(-x^{2})^{D/2}}\delta_{ab}\,,\label{eq:pert-prop-x} \\
   S_{ab}^{G}(x)= &\frac{i}{32\pi^{D/2}}\Gamma(\frac{D}{2}-1) \frac{\slashed{x}\sigma^{\mu\nu}+\sigma^{\mu\nu}\slashed{x}}{(-x^{2})^{D/2-1}}T^{r}_{ab}{:}g_{s}G_{\mu\nu}^{r}{:}\,,\label{eq:G-prop-x}
\end{align}
where $\slashed{x}\equiv \gamma_{\mu}x^{\mu}$, $g_{s}$ is the strong coupling constant, $G_{\mu\nu}^{r}$ is the background gluon field strength tensor, $T^{r}_{ab}$ are the generators of the SU(3) color group, $a$, $b$ are the color indices, and $\sigma^{\mu\nu}=\frac{i}{2}[\gamma^{\mu},\gamma^{\nu}]$. The colons denote the normal ordering of the operator, which will be sandwiched between the non-perturbative vacuum states to form the condensate when the OPE calculation is finished. Evidently, these propagators remain finite as $D\rightarrow 4$, and they can be directly applied in light-quark QCD sum rule calculations.

However, when considering light-quark propagators with higher-dimensional background field operators, such as those involving a covariant derivative of a background gluon field strength tensor or multiple background gluon fields, massless-quark-induced IR divergences can emerge. Furthermore, their coordinate-space expressions become considerably more complicated than their momentum-space counterparts.

\begin{figure}[htbp]
    \centering
    \subfigure[]{
        \includegraphics[width=0.15\textwidth]{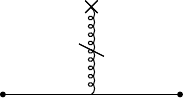}
        \label{fig1a}
        }\qquad
    \subfigure[]{
        \includegraphics[height=0.09\textwidth]{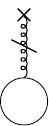}
        \label{fig1b}
        }
    \caption{Diagram (a) denotes the propagator with a covariant derivative of a background gluon field strength tensor $G^{r}_{\mu\nu}$. The slash mark on the gluon line represents a covariant derivative acting on the gluon field strength tensor. Diagram (b) represents the one-loop contribution of ${:}\chi_{a}(0)\overline{\chi}_{b}(0){:}$, which is treated as the IR counterterm for Diagram (a).} 
    \label{fig: gluon_propagator}
\end{figure}

As an example, the propagator involving a covariant derivative of a background gluon field strength tensor (shown in Figure~\ref{fig1a}) reads in momentum space~\cite{Reinders:1984sr}
\begin{equation}
    S_{ab}^{DG}(p)=\frac{i}{3}\frac{1}{\slashed{p}}\left(f_{\nu\mu\rho}+f_{\nu\rho\mu}\right)\frac{1}{\slashed{p}}T^{r}_{ab} {:}g_{s}G_{\mu\nu;\,\rho}^{r}{:}\,,\label{eq:DG-prop-mom}
\end{equation}
where $f_{\nu\mu\rho}=\gamma_{\nu}\frac{1}{\slashed{p}}\gamma_{\mu}\frac{1}{\slashed{p}}\gamma_{\rho}$, $G_{\mu\nu;\,\rho}^{r}=\tilde{D}_{\rho}^{rs}G_{\mu\nu}^{s}$, $\tilde{D}_{\rho}^{rs}=\delta^{rs}\partial_{\rho}-g_{s}f^{rst}A_{\rho}^{t}$ is the covariant derivative operator in the adjoint representation and $A_{\rho}^{t}$ is the background gluon field. Note that the IR divergence in $S_{ab}^{DG}(p)$ is implicit in momentum space. Under Fourier transformation, the exponential factor $e^{-ip\cdot x}$ regulates the ultraviolet (UV) behavior but cannot regulate the IR behavior ($p\rightarrow 0$). Therefore, the IR divergence becomes explicit in coordinate space.

The corresponding coordinate-space expression obtained via Fourier transformation reads~\cite{Zhang:2025fuz}:
\begin{align}
    S_{ab}^{DG}(x)=&\frac{\Gamma(\frac{D}{2}-2)\left(\gamma^{\mu}\gamma^{\rho}\gamma^{\nu}+\gamma^{\rho}\gamma^{\mu}\gamma^{\nu}-4 g^{\mu\rho}\gamma^{\nu}\right)}{96\pi^{D/2}(-x^{2})^{D/2-2}}T^{r}_{ab}{:}g_{s}G_{\mu\nu;\,\rho}^{r}{:}  \nonumber \\
    &+\frac{\Gamma(\frac{D}{2}-1)(x^{\mu}\gamma^{\rho}\slashed{x}\gamma^{\nu}+x^{\rho}\gamma^{\mu}\slashed{x}\gamma^{\nu})}{48\pi^{D/2}(-x^{2})^{D/2-1}}T^{r}_{ab}{:} g_{s}G_{\mu\nu;\,\rho}^{r}{:} \,.\label{eq:DG-prop-x}
\end{align}
The first term on the right-hand side of Eq.~\eqref{eq:DG-prop-x} exhibits a divergence as $D\rightarrow 4$ due to the factor $\Gamma(\frac{D}{2}-2)$. This is an IR divergence arising from the massless limit for the light quarks. In dimensional regularization with $D=4-2\epsilon$, such an IR divergence appears as a pole in $1/\epsilon$, which corresponds to the logarithmic mass singularity $\ln(m_{q}^{2})$ encountered in the heavy-quark expansion technique~\cite{Bagan:1985zp,Broadhurst:1984rr,Broadhurst:1985js,Grozin:1986xh,Grozin:1994hd}.

When propagators linear in the quark mass are considered, as commonly employed for strange quarks, the same IR divergences can appear. For example, the coordinate-space propagator involving a background gluon field strength tensor and a quark mass term reads
\begin{align}
    S_{ab}^{m_{q}G}(x)=&\frac{\Gamma(\frac{D}{2}-2)}{32\pi^{D/2}}\frac{{:}g_{s}m_{q}G_{\mu\nu}^{r}{:}T^{r}_{ab}\sigma^{\mu\nu}}{(-x^{2})^{D/2-2}} \,.\label{eq:mG-prop-x}
\end{align}
Again, the factor $\Gamma(\frac{D}{2}-2)$ in Eq.~\eqref{eq:mG-prop-x} produces an IR divergence in the limit $D\rightarrow 4$. We can expand Eq.~\eqref{eq:mG-prop-x} in powers of $\epsilon$ to expose this divergence explicitly:
\begin{align}
    S_{ab}^{m_{q}G}(x)=&-\frac{{:}g_{s}m_{q}G_{\mu\nu}^{r}{:}T^{r}_{ab}\sigma^{\mu\nu}}{32\pi^2\epsilon}\nonumber\\
    &-\frac{{:}g_{s}m_{q}G_{\mu\nu}^{r}{:}T^{r}_{ab}\sigma^{\mu\nu}}{32\pi^2}\left(\ln(-x^{2})+\gamma_{E}+\ln(\pi) \right)+O(\epsilon)\,,\label{eq:mG-prop-x-expand}
\end{align}
where $\gamma_{E}$ is the Euler constant. 

However, one cannot simply expand the propagators in $\epsilon$ and discard $O(\epsilon)$ terms at the outset, because such terms can enter loop integrals in OPE calculations and contribute finite parts to the Wilson coefficients. Therefore, it is necessary to keep the full $D$-dimensional expressions of the propagators until the final step of the OPE calculation. 

The presence of IR divergences in these propagators indicates that using them directly in OPE calculations will produce IR-divergent Wilson coefficients. As we will show in the next section, a subtraction procedure can remove these divergences at the propagator level beforehand.

\section{The cancellation of IR divergences}\label{sec:3}
In QCD sum rules, the background quark field $\chi_{a}(x)$ and the background gluon field strength tensor $G_{\mu\nu}^{r}(x)$ are employed to describe the non-perturbative effects of the QCD vacuum. The propagator for the background quark field is expressed as
\begin{align}
   Q_{ab}(x)={:}\chi_{a}(x)\overline{\chi}_{b}(0){:}\,.\label{eq:Q-prop-x}
\end{align}
Using the fixed-point (Fock-Schwinger) gauge $x^{\mu}A_{\mu}^{a}(x)=0$, $Q_{ab}(x)$ can be expanded at $x=0$, 
\begin{align}
     Q_{ab}(x)=&\sum_{n = 0}^{\infty}\frac{1}{n!}x^{\alpha_{1}}\cdots x^{\alpha_{n}}{:}(D_{\alpha_{1}}\cdots D_{\alpha_{n}}\chi_{a})(0)\overline{\chi}_{b}(0){:}\nonumber\\
     =&\sum_{n = 0}^{\infty}\frac{1}{n!}x^{\alpha_{1}}\cdots x^{\alpha_{n}}Q_{ab,\{\alpha_{1},\ldots,\alpha_{n}\}}^{(n)}(0)\,,\label{eq:Q-prop-expand}
\end{align}
where $D_{\alpha_{n}}=\partial_{\alpha_{n}}-ig_{s}A_{\alpha_{n}}$ denotes the covariant derivative and $Q_{ab,\{\alpha_{1},\ldots,\alpha_{n}\}}^{(n)}(0)={:}(D_{\alpha_{1}}\cdots D_{\alpha_{n}}\chi_{a})(0)\overline{\chi}_{b}(0){:}$ are operators which can either be directly vacuum-sandwiched to form the quark condensates, or combined with the background gluon field and then vacuum-sandwiched to form the hybrid condensates.

In the background field method, the QCD Lagrangian includes the interaction term between the background quark and gluon fields, ${:}g_{s}\overline{\chi}\slashed{A}\chi${:}~\cite{Reinders:1984sr,Govaerts:1983ka,Govaerts:1984bk}. Inserting this interaction vertex one or more times into $Q_{ab,\{\alpha_{1},\ldots,\alpha_{n}\}}^{(n)}(0)$ and contracting the background quark fields generates a one-loop contribution. Through the Fock-Schwinger gauge expansion, this contribution is expressed in terms of the background gluon field strength tensor operators of the same dimension as $Q_{ab,\{\alpha_{1},\ldots,\alpha_{n}\}}^{(n)}(0)$. As we will see, this one-loop contribution plays a key role in eliminating the IR divergences.

To illustrate the removal of the IR divergence in the light-quark propagator, we first set the quark mass $m_q$ to zero and consider $Q_{ab}^{(0)}(0)={:}\chi_{a}(0)\overline{\chi}_{b}(0){:}$. Inserting the interaction vertex into $Q_{ab}^{(0)}(0)$ generates the one-loop contribution shown in Figure~\ref{fig1b}. Since the dimension of $Q_{ab}^{(0)}(0)$ is three and there is no other quantity to supply the dimension, the one-loop contribution of $Q_{ab}^{(0)}(0)$ is expressed only in terms of the background gluon field strength tensor operator of dimension three, which is ${:}g_{s}G_{\mu\nu;\,\rho}^{r}{:}$. 

Since the background fields are soft~\cite{Shifman:1978bx,Shifman:1978by,Reinders:1984sr,Grozin:1986xh}, the one-loop contribution of $Q_{ab}^{(0)}(0)$ is dominated by the low-momentum region ($p\sim 0$) and produces an IR divergence. This one-loop contribution thus serves as the IR counterterm for $S_{ab}^{DG}(x)$ and is given by 
\begin{align}
   I^{DG}_{ab}=& {:}g_{s}G_{\mu\nu;\,\rho}^{r}{:} \int\limits_{p\sim 0} \frac{d^{D}p}{(2\pi)^{D}}\frac{i}{3}\frac{1}{\slashed{p}}\left(f_{\nu\mu\rho}+f_{\nu\rho\mu}\right)\frac{1}{\slashed{p}}T^{r}_{ab}\,,\label{eq:DG-counterterm-1}
\end{align}
where the integrand takes the same form as $S_{ab}^{DG}(p)$ in Eq.~\eqref{eq:DG-prop-mom}. If the integration is extended to the full momentum range, the loop integral in Eq.~\eqref{eq:DG-counterterm-1} becomes a scaleless integral. In dimensional regularization, the scaleless integral vanishes because the UV and IR divergences cancel each other~\cite{Collins:1984xc,Smirnov:2012gma}. Therefore, there is no UV divergence in Eq.~\eqref{eq:DG-counterterm-1} and only the IR divergence remains. 

To calculate $I^{DG}_{ab}$, we exploit the fact that the UV and IR divergences are equal and opposite in the scaleless integral. We extract the UV part using the \texttt{PaVeUVpart} function~\cite{Feng:2012kd,Sulyok:2006prd} in FeynCalc~\cite{Shtabovenko:2023idz}, and take the opposite sign to obtain the IR part. The result is
 \begin{align}
    I^{DG}_{ab}&=-\frac{\Gamma(2-\frac{D}{2})}{3\times(4\pi)^{D/2}}\gamma^{\mu}g^{\nu\rho}T^{r}_{ab}{:}g_{s}G_{\mu\nu;\,\rho}^{r}{:} \nonumber\\
    &=Z^{DG,\,\mu\nu\rho,\, r}_{ab}{:}g_{s}G_{\mu\nu;\,\rho}^{r}{:}\,,\label{eq:DG-counterterm-2}
 \end{align} 
where 
\begin{align}
    Z^{DG,\,\mu\nu\rho,\, r}_{ab}=-\frac{\Gamma(2-\frac{D}{2})}{3\times(4\pi)^{D/2}}\gamma^{\mu}g^{\nu\rho}T^{r}_{ab}\,.\label{eq:DG-counterterm-Z}
\end{align}
Here, we have adopted the $\overline{MS}$ scheme such that $\frac{\pi^{2}}{4-D}\rightarrow\frac{1}{2\times(4\pi)^{D/2}}\Gamma(2-\frac{D}{2})$. We keep the result in $D$ dimensions since the $\epsilon$ expansion should be performed at the last step of the OPE calculation. If we subtract $I^{DG}_{ab}$ from $S_{ab}^{DG}(x)$, we obtain an IR-finite propagator $F_{ab}^{DG}(x)$ as $D\rightarrow 4$,
\begin{align}
    F_{ab}^{DG}(x)=&S_{ab}^{DG}(x)-I^{DG}_{ab}\nonumber\\
    =&-\frac{{:}g_{s}G_{\mu\nu;\,\rho}^{r}{:} T^{r}_{ab}}{48\pi^2}\left(\frac{x^{\mu}\gamma^{\rho}\slashed{x}\gamma^{\nu}+x^{\rho}\gamma^{\mu}\slashed{x}\gamma^{\nu}}{x^{2}}\right.\nonumber\\
    &+\gamma^{\mu}g^{\nu\rho}\left(\ln(-x^{2})+2\gamma_{E}-2\ln(2)\right)\biggr)+O(\epsilon)\,.\label{eq:DG-prop-x-finite}
\end{align}

If the quark mass term is retained to linear order in $m_{q}$, the one-loop contributions of $Q_{ab}^{(0)}(0)$ will also contain a term proportional to ${:}g_{s}m_{q}G_{\mu\nu}^{r}{:}$, whose total dimension is also three. This one-loop contribution is given by
\begin{align}
    I^{m_{q}G}_{ab}=& {:}g_{s}m_{q}G_{\mu\nu}^{r}{:} \int\limits_{p\sim 0} \frac{d^{D}p}{(2\pi)^{D}}\frac{-i}{2}\frac{\sigma^{\mu\nu}}{p^{4}}T^{r}_{ab}\nonumber\\
    =&-\frac{\Gamma(2-\frac{D}{2})}{2\times(4\pi)^{D/2}}\sigma^{\mu\nu}T^{r}_{ab}{:}g_{s}m_{q}G_{\mu\nu}^{r}{:} \nonumber\\
    =&Z^{m_{q}G,\,\mu\nu,\, r}_{ab}{:}g_{s}m_{q}G_{\mu\nu}^{r}{:} \,,\label{eq:mG-counterterm}
\end{align}
where
\begin{align}
    Z^{m_{q}G,\,\mu\nu,\, r}_{ab}=-\frac{\Gamma(2-\frac{D}{2})}{2\times(4\pi)^{D/2}}\sigma^{\mu\nu}T^{r}_{ab}\,.\label{eq:mG-counterterm-Z}
\end{align}
Again, we subtract $I^{m_{q}G}_{ab}$ from $S_{ab}^{m_{q}G}(x)$, and the resulting expression is finite as $D\rightarrow 4$,
\begin{align}
    F_{ab}^{m_{q}G}(x)=&S_{ab}^{m_{q}G}(x)-I^{m_{q}G}_{ab}\nonumber\\
    =&-\frac{{:}g_{s}m_{q}G_{\mu\nu}^{r}{:} T^{r}_{ab}}{32\pi^2}\sigma^{\mu\nu}\nonumber\\
    &\times\left(\ln(-x^{2})+2\gamma_{E}-2\ln(2)\right)+O(\epsilon)\,.\label{eq:mG-prop-x-finite}
\end{align}
We note that the finite part of Eq.~\eqref{eq:mG-prop-x-finite} coincides with the result in Ref.~\cite{Pasupathy:1986pw}, where an IR regulator $\Lambda$ is introduced. 

The examples above demonstrate that the one-loop contributions of $Q_{ab}^{(0)}(0)$ indeed serve as counterterms to eliminate the IR divergences in $S_{ab}^{DG}(x)$ and $S_{ab}^{m_{q}G}(x)$. The resulting expressions are finite as $D\rightarrow 4$ and can be safely used in the OPE calculation. Therefore, it is natural to extend the IR-divergence subtraction method to the general case. 

For $Q_{ab,\{\alpha_{1},\ldots,\alpha_{n}\}}^{(n)}(0)$, the one-loop contributions comprise terms involving various background field operators of the same dimension as $Q_{ab,\{\alpha_{1},\ldots,\alpha_{n}\}}^{(n)}(0)$. The background field operator $O$ is built from the background gluon field strength tensor $G_{\mu\nu}^{r}$, its covariant derivatives and the quark mass, such as ${:}g_{s}m_{q}G_{\mu\nu}^{r}{:}$ and ${:}g_{s}G_{\mu\nu;\,\rho}^{r}{:}$. 

The one-loop contribution of $Q_{ab,\{\alpha_{1},\ldots,\alpha_{n}\}}^{(n)}(0)$ corresponding to a specific background field operator $O$ is given by
\begin{align}
    I_{ab,\{\alpha_{1},\ldots,\alpha_{n}\}}^{O}=&\int\limits_{p\sim 0} \frac{d^{D}p}{(2\pi)^{D}}(-i p_{\alpha_{1}})\cdots (-i p_{\alpha_{n}})S_{ab}^{O}(p)\nonumber\\
    =&Z^{O}_{ab,\,\{\alpha_{1},\ldots,\alpha_{n}\}}O\,,\label{eq:Q-counterterm-1}
\end{align} 
where $S_{ab}^{O}(p)$ is obtained by contracting the background quark fields with the interaction vertices to generate the operator $O$. Since the interaction between the background quark and gluon fields takes the same form as the usual quark-gluon interaction, $S_{ab}^{O}(p)$ has the same form as the corresponding momentum-space quark propagator in QCD sum rules, and we adopt the same notation. The factor $(-i p_{\alpha_{1}})\cdots (-i p_{\alpha_{n}})$ in the integrand of Eq.~\eqref{eq:Q-counterterm-1} originates from the covariant derivatives in $Q_{ab,\{\alpha_{1},\ldots,\alpha_{n}\}}^{(n)}(0)$. $Z^{O}_{ab,\,\{\alpha_{1},\ldots,\alpha_{n}\}}$ is a dimensionless coefficient that contains the Dirac matrix structure and the IR divergence of the loop integral associated with $O$. The integral $I_{ab,\{\alpha_{1},\ldots,\alpha_{n}\}}^{O}$ has dimension $3+n$; it should be multiplied by the factor $\frac{1}{n!}x^{\alpha_{1}}\cdots x^{\alpha_{n}}$ from the expansion of $Q_{ab}(x)$ to form the dimension-three counterterm for $S_{ab}^{O}(x)$,
\begin{align}
    R^{O}_{ab}(x)=&\frac{1}{n!}x^{\alpha_{1}}\cdots x^{\alpha_{n}} I_{ab,\{\alpha_{1},\ldots,\alpha_{n}\}}^{O}\,.\label{eq:Q-counterterm-2}
\end{align}
In the $Q_{ab}^{(0)}(0)$ case, $R^{DG}_{ab}(x)=I^{DG}_{ab}$ and $R^{m_{q}G}_{ab}(x)=I^{m_{q}G}_{ab}$. Thus, the IR-finite propagator $F^{O}_{ab}(x)$ corresponding to $S_{ab}^{O}(x)$ is obtained by subtracting $R^{O}_{ab}(x)$ from $S_{ab}^{O}(x)$,
\begin{align}
    F^{O}_{ab}(x)=&S_{ab}^{O}(x)-R^{O}_{ab}(x)\,.\label{eq:prop-x-finite}
\end{align}

Unlike the original propagator $S_{ab}^{O}(x)$, the IR-finite propagator $F^{O}_{ab}(x)$ is free of $\frac{1}{\epsilon}$ IR poles. Therefore, in practical QCD sum rule calculations, one can replace $S_{ab}^{O}(x)$ with $F^{O}_{ab}(x)$ and proceed with the standard OPE procedure without needing to consider condensate mixing. 

The formula in Eq.~\eqref{eq:prop-x-finite} has been applied to eliminate the IR divergences in the dimension-eight four-gluon condensates for the light hybrid baryons~\cite{Wang:2024lnv}. In the resulting Wilson coefficient, a double-logarithm term $\ln^{2}\left(\frac{-q^{2}}{\mu^{2}}\right)$ appears, which yields a single-logarithm term $\ln\left(\frac{s}{\mu^{2}}\right)$ in the spectral density. The results have been cross-checked using the methods of Refs.~\cite{Generalis:1983hb,Broadhurst:1984rr} and the coefficients in Ref.~\cite{Broadhurst:1985js}, and they are found to be consistent with each other.

To further illustrate the importance of the IR subtraction in obtaining the 
correct OPE result, we consider the $m_{s}\qGqb$ contribution arising from 
$S^{m_{q}G}_{ab}(x)$ for the strange quark in our previous 
work~\cite{Zhang:2025fuz}. Without the subtraction, the correlation 
function for the current $P_{6}$ in Ref.~\cite{Zhang:2025fuz} receives a 
contribution of the form
\begin{align}
    \Pi^{m_{s}\qGqb}(q^{2})=&\frac{q^{2}}{256\pi^{4}\epsilon}m_{s}\qGqb\nonumber\\ 
    &-m_{s}\qGqb\frac{q^{2}}{1536 \pi ^4}\left(12 \ln \left(-q^2\right)-17\right.\nonumber\\
    &\left.+12\Big(\gamma_{E}-\ln (4\pi)\Big)\right)+O(\epsilon)\,,\label{eq:P6-mG-contribution}
\end{align}
where the logarithmic term $\ln(-q^{2})$ contributes to the spectral 
density. Including the counterterm $R^{m_{q}G}_{ab}(x)$ in the OPE 
calculation yields
\begin{align}
    \Pi^{m_{s}\qGqb}_{CT}(q^{2})=&\frac{q^2}{128 \pi^{4} \epsilon }m_{s}\qGqb\nonumber\\
    &-m_{s}\qGqb\frac{q^{2} }{384 \pi ^4} \left(3 \ln \left(-q^2\right)-1\right. \nonumber\\
    &\left.+6\Big(\gamma_{E} -\ln (4\pi )\Big)\right)+O(\epsilon)\,,\label{eq:P6-mG-counterterm-contribution}
\end{align}
where the subscript ``CT'' denotes the counterterm contribution. Subtracting $\Pi^{m_{s}\qGqb}_{CT}(q^{2})$ from $\Pi^{m_{s}\qGqb}(q^{2})$ gives
\begin{align}
    \Pi^{m_{s}\qGqb}_{\text{IR-subtracted}}(q^{2})=&\Pi^{m_{s}\qGqb}(q^{2})-\Pi^{m_{s}\qGqb}_{CT}(q^{2})\nonumber\\
    =&-\frac{q^2}{256 \pi^{4} \epsilon }m_{s}\qGqb+\frac{13q^2}{1536\pi^{4}}m_{s}\qGqb\nonumber\\
    &+\frac{q^2(\gamma_{E} -\ln (4\pi ))}{128 \pi^{4}}m_{s}\qGqb+O(\epsilon)\,,\label{eq:P6-mG-subtracted-contribution}
\end{align}
which is free of $\ln(-q^{2})$ and therefore does not contribute to the 
spectral density. The remaining $\frac{1}{\epsilon}$ pole in the above is of UV origin and will be removed by renormalization. This demonstrates that the IR subtraction is essential for obtaining the correct OPE result: without it, the spectral density would contain spurious contributions from IR-divergent propagators.

\section{Discussion and conclusions}\label{sec:4}
In the two methods proposed in Refs.~\cite{Generalis:1983hb,Broadhurst:1984rr}, the key to the cancellation of IR divergences lies in the mixing coefficients between quark and gluon condensates of the same dimension. Specifically, inserting the interaction term from the QCD Lagrangian into the quark condensates annihilates the low-momentum background quark fields and excites soft background gluon fields. This process effectively expresses the quark condensates in terms of the corresponding gluon condensates. Schematically, this mixing is represented by a one-loop diagram, and evaluating this diagram yields the mixing coefficient. By multiplying this mixing coefficient by the original Wilson coefficient and subtracting it from the corresponding Wilson coefficient, the IR divergence is removed. This is the mechanism for the cancellation of IR divergences in these two methods.

As we have shown in Section~\ref{sec:2}, the IR divergences appearing in the Wilson coefficients of the gluon condensates originate from the absence of an IR regulator for the light-quark propagator, which causes the coordinate-space propagator to become explicitly divergent as $D\rightarrow 4$ in dimensional regularization. Rather than removing these divergences at the OPE level through condensate mixing, our method operates directly at the propagator level: we insert the interaction term into the expansion of the background quark propagator $Q_{ab}(x)$ at $x=0$ to generate a one-loop contribution expressed in terms of the background field operator $O$. The coefficient $Z^{O}_{ab,\,\{\alpha_{1},\ldots,\alpha_{n}\}}$ obtained from this one-loop diagram plays the same role as the mixing coefficient in the previous methods. Subtracting $R^{O}_{ab}(x)$ from the original propagator $S_{ab}^{O}(x)$ removes the IR divergences at the propagator level, yielding the IR-finite propagator $F^{O}_{ab}(x)$. It is worth stressing that this subtraction not only removes the $\frac{1}{\epsilon}$ IR poles but is also essential for obtaining the correct spectral density.

In conclusion, the subtraction formula in Eq.~\eqref{eq:prop-x-finite} provides an improved method for eliminating IR divergences at the propagator level in light-quark QCD sum rules. This simplifies the computation compared to the previous approaches, which remove IR divergences at the OPE level. Moreover, it takes a form similar to the subtraction of UV divergences in the renormalization procedure, making it more intuitive and easier to understand. This method can be readily applied to QCD sum rule studies involving light quarks, such as exotic hadron spectroscopy, where high-dimensional condensates are frequently encountered.

\begin{acknowledgments}
This work is supported by the National Natural Science Foundation of China under Grant Nos. 12075019, 12575153, and 12221005, and by the Chinese Academy of Sciences under Grant No. YSBR-101. QNW is supported by the Natural Science Foundation of Liaoning Province of China under Grant No. 2025-BS-0816 and the Doctoral Startup Project of Bohai University under Grant No. 0525bs003. 
\end{acknowledgments}

%

\end{document}